\documentclass[a4paper,12pt]{article}

\usepackage{ifpdf}

\newif\ifpdf
\ifx\pdfoutput\undefined
  \pdffalse
\else
  \pdfoutput=1
  \pdftrue
\fi

\RequirePackage{xspace} %
\RequirePackage{subfigure} %
\RequirePackage[centertags]{amsmath} %
\RequirePackage{amssymb}
\RequirePackage{wrapfig} %
\RequirePackage{calc} %
\RequirePackage{ifthen}
\RequirePackage{tabularx} %
\RequirePackage{flafter} %
\RequirePackage{fancyhdr} %

\ifpdf
  \RequirePackage[pdftex]{color}%
  \RequirePackage{colortbl}%
  \RequirePackage{array}%
  \RequirePackage[pdftex]{graphicx}

  \RequirePackage[ pdftex, plainpages = false, pdfpagelabels,
                 pdfpagelayout = useoutlines,
                 bookmarks,
                 breaklinks = true,
                 linktocpage,
                 pagebackref,                      % to include page numbers in bibliography
                 colorlinks = true,
                 linkcolor = blue,
                 urlcolor  = blue,
                 citecolor = blue,
                 anchorcolor = blue,
                 hyperindex = true,
                 hyperfigures
                 ]{hyperref}

\else
  \RequirePackage{color}
  \RequirePackage{colortbl}
   \RequirePackage{array}
  \RequirePackage[dvips]{graphicx}
  \RequirePackage{hyperref}
  \usepackage{rotating}
\fi

\usepackage{makeidx} % enable indexing
\usepackage{setspace} %enable setting spacing, %e.g. singlespace, onehalfspace, and doublespace
\usepackage{rotating} %enable sidewaysfigure
\usepackage{ecltree}
\usepackage{epic}
\usepackage{supertabular}  % this will allow the table not to break
\usepackage{color}
\usepackage{exscale}
\usepackage{fontenc}
\usepackage{ifthen}
\usepackage{latexsym}
\usepackage{makeidx}
\usepackage{syntonly}
\usepackage{inputenc}
\usepackage{graphicx}
\usepackage{setspace}
\usepackage{caption2}
\usepackage[english]{babel}
\usepackage[square, comma,numbers,sort&compress]{natbib}
\usepackage{hypernat}
\usepackage{boxedminipage}
\usepackage{framed}
\usepackage{longtable}
\usepackage[all]{hypcap}    %included to make the hyperlink go to the top of figure
\usepackage{algorithm2e}
\usepackage{algorithmic}
\usepackage{lscape}
\usepackage{pdflscape}

\newcommand{\etal}{{\it et al}}

\newcommand{\note}[1]{\marginpar[left]{\singlespace \tiny #1}}
\newcommand{\pois}{Poiseuille}
\newcommand{\HB}{Herschel-Bulkley}

\newcommand{\sS}     {\tau}           % Shear stress
\newcommand{\wsS}    {\sS_{w}}        % Wall shear stress
\newcommand{\ysS}    {\sS_{o}}        % Yield stress
\newcommand{\typ}    {threshold yield pressure}
\newcommand{\ys}    {yield-stress}

\renewcommand{\sectionmark}[1]%
      {\markright{\thesection\ #1}} %stops it capitalizing. #1 has value of section name

\renewcommand{\note}[1]{}

\doublespace %\onehalfspace

\title
{ %
\vspace*{3.0cm} \LARGE{\bf Yield and Solidification of Yield-Stress Materials in Rigid Networks and Porous Structures} \vspace*{2.0cm} \\
}

\author{Taha Sochi\footnote{University College London, Department of Physics \& Astronomy, Gower Street, London, WC1E 6BT.
Email: t.sochi@ucl.ac.uk.} \vspace*{4.0cm}}

\begin{document}

\maketitle %
\pagenumbering{arabic}

\newpage
\phantomsection \addcontentsline{toc}{section}{Contents} %
\tableofcontents

%\newpage
%\phantomsection \addcontentsline{toc}{section}{List of Figures} %
%\listoffigures
%
%\phantomsection \addcontentsline{toc}{section}{List of Tables} %
%\listoftables

\newpage
\phantomsection \addcontentsline{toc}{section}{Abstract} \noindent
{\noindent \LARGE \bf Abstract} \vspace{0.5cm}\\
\noindent %

In this paper, we address the issue of threshold yield pressure of yield-stress materials in rigid
networks of interconnected conduits and porous structures subject to a pressure gradient. We
compare the results as obtained dynamically from solving the pressure field to those obtained
statically from tracing the path of the minimum sum of threshold yield pressures of the individual
conduits by using the threshold path algorithms. We refute criticisms directed recently to our
previous findings that the pressure field solution generally produces a higher threshold yield
pressure than the one obtained by the threshold path algorithms. Issues related to the
solidification of yield stress materials in their transition from fluid phase to solid state have
also been investigated and assessed as part of the investigation of the yield point.


%%%%%%%%%%%%%%%%%%%%%%%%%%%%%%%%%%%  Head style  %%%%%%%%%%%%%%%%%%%%%%%%%%%%%%%%%%%
\pagestyle{headings} %
\addtolength{\headheight}{+1.6pt}
\lhead[{Chapter \thechapter \thepage}]%
      {{\bfseries\rightmark}}
\rhead[{\bfseries\leftmark}]%
     {{\bfseries\thepage}} %tell it to put page number at rhead
\headsep = 1.0cm               % Added 07 Sep 2006
%%%%%%%%%%%%%%%%%%%%%%%%%%%%%%%%%%%%%%%%%%%%%%%%%%%%%%%%%%%%%%%%%%%%%%%%%%%%%%%%%%%%

\newpage
%XXXXXXXXXXXXXXXXXXXXXXXXXXXXXXXXXXXXXXXXXXXXXXXXXXXXXXXXXXXXXXXXX
\section{Introduction}

Yield-stress materials are commonplace in nature and industry. They include very common biological
fluids like blood \cite{SochiNonNewtBlood2013} as well as many polymeric solutions used in
reservoir engineering and pharmaceutical manufacturing. These materials are characterized by
behaving like solids below a certain threshold stress and like fluids above. There are many
controversies about the nature of these materials, their rheological definition, and even their
bare existence. They are problematic both experimentally, as they behave strangely and sometimes
unpredictably, and theoretically as they are difficult to model and simulate computationally. There
are several rheological models that have been proposed for modeling these materials; some of the
most common ones are Bingham, \HB, and Casson. However, almost all the available rheological models
that characterize the \ys\ behavior are empirical in essence and phenomenological in nature
\cite{Skellandbook1967, BirdbookAH1987, BarnesbookHW1993, CarreaubookKC1997, Barnes1999}.

The above-mentioned problems that associate the bulk rheology of \ys\ materials are aggravated by
more complications and difficulties when their rheology in porous structures and networks of
interconnected conduits is investigated experimentally or theoretically. Several fluid-structure
interaction factors emerge in such situations to play intricate defining roles in the overall
conduct of such systems. For instance, the effect of tortuosity and shape irregularities of the
conduits inside such structures makes the local yield point highly dependent on several geometric
and topological factors that are difficult to predict and model \cite{SochiYield2010,
SochiFeature2010}.

In the mobilization of \ys\ materials through networks of interconnected conduits and porous
structures, there is an important issue about how to predict the \typ\ of such materials saturating
such media. In this regard, there are two main approaches to predict the yield point: (a)
determining the \typ\ dynamically by finding the pressure field which is normally obtained through
solving the balance equations of the flow system that are based on the conservation principles and
constitutive fluid models, and (b) determining the \typ\ statically through using the threshold
path algorithms such as the invasion percolation with memory \cite{KharabafY1997} and the path of
minimum pressure \cite{SochiThesis2007, SochiB2008, SochiYield2010} which trace the route that
minimizes the sum of the \typ s of the route conduits inside these structures. These algorithms are
based on the inert geometry of the individual conduits and the rheology of the \ys\ materials
without the involvement of dynamic factors.

Sochi and Blunt \cite{SochiB2008} and Sochi \cite{SochiYield2010} (henceforth these references are
called SB) have investigated this problem and concluded that the \typ\ obtained from solving the
flow system is generally higher than the one obtained from the threshold path algorithms. This was
justified by several factors; the main ones are (a) the rejection of the underlying assumption of
the threshold path algorithms that the \typ\ of serially connected conduits is equal to the sum of
their individual \typ s, (b) dynamic factors related to obtaining a stable and consistent pressure
field configuration, (c) the effect of the tortuosity on the pressure field and its direct impact
on the nodal pressure of the intermediate nodes and hence the yield point of the threshold path,
and (d) the communication of these intermediate nodes with the global pressure field through
conduits connected to these nodes but are not part of the threshold path.

Recently, Balhoff \etal\ \cite{BalhoffRKMP2012} (henceforth called BRKMP) conducted a study in
which they investigated this issue, among other issues, in detail and challenged the previous
findings of SB. They argued that the \typ\ obtained from solving the balance equations must be the
same as the one obtained from the threshold path algorithms. They supported their theoretical
reasoning by flow simulations in which they used a robust solving scheme based on the
Newton-Raphson method in conjunction with the mass conservation and characteristic flow models.
They even produced a mathematical proof using a graph theory framework to back their findings.

In this context, we should distinguish between two transition points for yield stress materials
between the solid state and the fluid phase. The first one, which we call the yield point, is the
transition from the solid-like state to the fluid state; and the second one, which we call the
solidification or blockage point, is the transition from the fluid state to the solid state. These
two points are in general different due to the effect of the initial flow conditions and hysteresis
and hence the experimental and computational searching techniques for these two points should be
different as well. However, the two problems are closely linked although there seems to be little
interest in the solidification problem due, apparently, to a common belief that the two points are
the same. We will discuss the solidification point as part of our investigation to the yield point
but we will not go deep into this investigation due to the specific objectives of the current
study.

In our view, the yield point should be identified by a gradual and continuous increase in the
pressure drop, whether across the bulk material or single conduit or interconnected structure of
multiple conduits, on starting from a confirmed solid state point such as zero pressure drop, while
the solidification point should be identified by a gradual and continuous decrease in the pressure
drop on starting from a well established fluid state point. There is also the possibility of a
sudden and non-continuous change in the pressure drop on a \ys\ material in its solid or fluid
state which may or may not result in a transition in the state of material. The latter possibility
is relevant to identifying the yield point if the material was initially a solid with a sudden
pressure increase and to identifying the solidification point if the material was initially a fluid
with a sudden pressure decrease. Other possibilities can also be considered but they are of little
relevance to the current investigation and hence they will be ignored.

Which method, gradual or sudden, should be used to determine the yield and solidification points is
a matter of convention as long as the conditions are stated unambiguously. However, it is very
possible that the yield and solidification points obtained from a sudden change in the pressure
drop are not the same as the ones obtained from a gradual change. One potential reason for this is
transitional instabilities although other reasons are also possible. In this case, more than one
point for yield and solidification, which depend on the pressure application method, should be
accepted if it is supported by experimental evidence. More discussion about these issues can be
found in \cite{SochiYield2010} and section \ref{Discussion}.

In the present paper, we discuss the issue of yield point in detail and challenge the findings of
BRKMP. Our main objection to the BRKMP criticism and findings is that what they claim to be the
yield point is in fact more appropriate to be the solidification point of the \ys\ materials on a
gradual decrease of the pressure from above the sum of thresholds where BRKMP assumed the material
has already yielded and hence it is a fluid, to the point of blockage where the system converges to
the solid state because it has reached the sum of the threshold yield pressures on the threshold
path. We also present two mathematical proofs for our proposal that the \typ\ of an ensemble of
serially-connected conduits is in general greater than the sum of their individual \typ s. One of
these proofs is based on the assumption that \ys\ materials prior to reaching their yield point are
fluids with very high viscosity, and the other proof is based on the assumption that \ys\ materials
are solids prior to yield. A mathematical argument has also been presented to show that finding a
mass-conserving consistent pressure field in a fluid-filled ensemble of interconnected conduits is
always possible for any type of fluid above its minimum mobilization pressure in the given
ensemble, where the minimum mobilization pressure is obtained from the sum of the minimum
mobilization pressures of the individual conduits in the ensemble. This argument is key to
identifying the circularity in the BRKMP argument.

The non-Newtonian fluids may be classified into two main categories, history-dependent which
include viscoelastic and thixotropic/rheopectic, and history-independent which are the purely
viscous non-Newtonian fluids that also include the Newtonian as a special case. The second category
may be equated with the generalized Newtonian fluids if yield-stress materials are accepted in this
category. Yield-stress can associate both history-dependent and history-independent fluids. For the
purpose of the present paper, these attributes are almost irrelevant as we are mainly interested in
the threshold yield point. Although history-dependent and history-independent attributes have very
strong impact on the flow, this is generally valid only above the very low-shear-rate regimes, i.e.
following yield and mobilization. The reason is that prior to yield any potential deformation is
minimal and hence any non-Newtonian effects, other than \ys, are negligible since the fluid is
still at its low-shear Newtonian plateau which characterizes almost all non-Newtonian fluids. The
low-shear Newtonian plateau can also be justified theoretically by the fact that all the
non-Newtonian rheological properties are strongly dependent on the rate of deformation, whether
shear or extension. We therefore do not differentiate in this paper between the history-dependent
and history-independent fluids as long as they are yield-stress materials, although we will
indicate the consequences of these properties on the \ys\ behavior briefly when necessary and where
relevant. However, history-dependent and history-independent attributes should have more
significant impact on the solidification point but in this study we investigate the solidification
point marginally as part of our investigation to the yield point.

We should also remark that in the present study we are only concerned with rigid networks and
porous structures; any tangible deformability, such as being elastic or viscoelastic, in these
structures requires further modeling considerations and hence complicates the modeling strategies
of the yield and solidification processes substantially.

%XXXXXXXXXXXXXXXXXXXXXXXXXXXXXXXXXXXXXXXXXXXXXXXXXXXXXXXXXXXXXXXXX
\section{Modeling Yield Stress in Porous Structures}

In this section we outline our proposals for modeling the mobilization of \ys\ materials in
networks of interconnected conduits and porous structures. We would like to insist that in the
present paper we consider these aspects from the viewpoint of threshold yield point only with
minimal consideration for the subsequent dynamic effects that automatically take place following
mobilization which will inevitably change the dynamics of the system, and hence the modeling
strategy, fundamentally. These effects must be taken into account thoroughly for a complete and
reliable \ys\ flow model. Also, there are still many detailed issues that should be dealt with at
the practical levels for implementing such models, such as convergence difficulties and convergence
enhancement techniques, which are very serious issues for such studies; but we do not consider
these issues here since they are out of the scope of the present paper. Some of these issues have
already been discussed in some of our previous papers (e.g. \cite{SochiTechnical1D2013,
SochiPoreScaleElastic2013}).

Although we presented a limited amount of computational work in this paper, due to its nature,
extensive computational work has been done in the background as part of this investigation to test
and verify various possibilities and aspects. We therefore feel obliged to provide a general
clarification about the computational framework which the current study relies upon. We have
already fully explained this framework in some of our previous publications and hence for the
purpose of saving space and avoiding repetition we refer the reader to the following papers:
\cite{SochiPoreScaleElastic2013, SochiConDivElastic2013, SochiCoDiNonNewt2013} where our
computational framework is fully explained. More relevant details may also be found in
\cite{SochiTechnical1D2013, SochiPois1DComp2013} although these are mainly related to a
one-dimensional finite element Newtonian flow model. We would also like to clarify that this
computational framework is different to the one used in our previous studies (e.g.
\cite{SochiThesis2007, SochiB2008, SochiVE2009, SochiYield2010}) and hence there is no ground for
potential criticisms based on the computational approach adopted in the previous studies. However,
we have no reason to believe that the previous results are incorrect or compromised because of the
previous computational framework and modeling strategies which, to the best of our knowledge, are
still valid in general.

The minimum pressure drop required to initiate the mobilization of a \ys\ material in its solid
state is called the \typ. The essential issue that determines the \typ\ of a \ys\ material
occupying a network of interconnected conduits or a porous structure is the pressure field
configuration inside such structures. Let us assume we have a solid porous structure filled with a
\ys\ material and we started from a zero pressure drop and kept varying the pressure drop across
the structure either gradually and continuously or through sudden changes. The crucial question
then is what is the pressure field configuration inside the porous structure as a function of the
applied pressure drop across the structure. If we can {\it a priori} determine the pressure field
spatially inside the structure as a function of the applied pressure drop for any given pressure
value then we can easily determine the threshold yield point by simply identifying the minimum
pressure drop across the structure that creates a path on which the pressure drop across each one
of its conduits exceeds the \typ\ for that conduit. For a perfectly circular cylindrical rigid tube
with a constant cross sectional area along its axial length, the \typ\ condition is given by
\cite{SochiThesis2007, SochiB2008, SochiYield2010}

\begin{equation}\label{yieldCondition}
\wsS = \ysS \hspace{1cm} \Longrightarrow \hspace{1cm} \Delta P_t = {\frac{2 L \ysS}{R}}
\end{equation}
where $\wsS$ is the shear stress at the tube wall, $\ysS$ is the \ys\ of the fluid, $\Delta P_t$ is
the pressure drop across the tube at the yield threshold, and $L$ and $R$ are the tube length and
radius respectively. It should be remarked that the condition given by Equation
\ref{yieldCondition} is a necessary but not sufficient condition for yield, as will be discussed
later.

It is noteworthy that our definition for the \typ\ of networks and porous structures and how it is
determined is based on some implicit assumptions about how the pressure field configuration inside
such structures changes in response to the applied pressure drop across these structures. Although
these assumptions are not self-evident, they seem to be generally accepted and hence we see no
necessity for discussing them in the present paper. However, there is one important assumption that
requires some clarification that is the assumption of continuity of yield above the \typ\ which,
although it seems intuitive, still requires a physical or mathematical justification. While we will
not discuss this issue here, we think the argument presented in Appendix \ref{AppA}, whose essence
is the possibility of finding a mass-conserving consistent pressure field for any type of fluid
above its minimum mobilization pressure, can be easily adjusted to provide such a proof. Due to the
fact that the flow rate is a strictly increasing and continuous function of pressure drop, the
yield condition of an ensemble, as soon as it is satisfied, will remain so on increasing the
pressure drop across the ensemble.

However, no one can completely rule out the possibility of a blockage subsequent to yield at a
pressure drop above the threshold yield pressure due to continuous or discrete transformations in
the system dynamics, especially with the involvement of complex non-Newtonian rheological factors
other than \ys, that change the pressure field configuration in a way that affects the yield
condition. This possibility may not be realistic in a simple one-dimensional network of
serially-connected conduits but it should be realistic for more complex two-dimensional and
three-dimensional networks and porous structures. Such a possibility should be seriously considered
for a complete \ys\ model which is out of the scope of the current paper as it is mainly focused on
the yield point succeeding a total blockage. Anyway, to avoid any possible disputes we could assume
that the previous statements in the former paragraphs related, explicitly or implicitly, to the
dependency of pressure field inside a structure on the pressure drop across it and subsequent
developments are just definitions or assumptions and hence they are part of our modeling strategy
and not considered as physical facts.

There are two main approaches for modeling \ys\ materials prior to reaching their yield point
whether in the bulk flow, single conduit flow or flow through networks of interconnected conduits
and porous structures; these two approaches are explained in the following subsections.

%XXXXXXXXXXXXXXXXXXXXXXX
\subsection{Highly-Viscous Fluid Approach}

According to this approach, the \ys\ materials prior to reaching their yield point are fluids with
very high viscosity. Therefore, they are distinguished by having a viscosity function whose
dependency on the shear stress is discontinuous at the yield point. Our modeling choice for the
highly-viscous fluid approach is to identify the pressure field prior to yield from solving the
balance equations assuming the fluid is Newtonian with a constant viscosity. As indicated early,
the Newtonian assumption prior to yield is very realistic one even for the highly non-Newtonian
fluids, because at these stages of negligible deformation, all non-Newtonian rheological effects,
except \ys, are absent as the fluid is still on its low-shear Newtonian plateau.

Therefore, to find the \typ\ we step up on the pressure ladder by starting from a confirmed
non-yield point and solve the pressure field at each pressure step using the \pois\ flow model. The
pressure field is then tested at each step to identify a possible inlet-to-outlet spanning path
whose all conduits have passed their threshold yield point, as given by Equation
\ref{yieldCondition} for cylindrical tubes. The minimum pressure drop that satisfies this condition
is taken to be the threshold yield pressure that defines the yield point. At and above this
pressure, the flow model for the mobilized parts will be subject to the adopted \ys\ rheological
fluid model such as \HB. Although the fluid prior to yield, according to this approach, is
theoretically assumed to be of very high-viscosity, computationally the value of the Newtonian
viscosity is irrelevant to the pressure field solution since the viscosity in the \pois\ model is a
conductance scale factor for the flow rate with no effect on the configuration of the pressure
field and hence any value for the viscosity will produce the same pressure field.

\begin{figure}[!h]
\centering{}
\includegraphics
[scale=0.45] {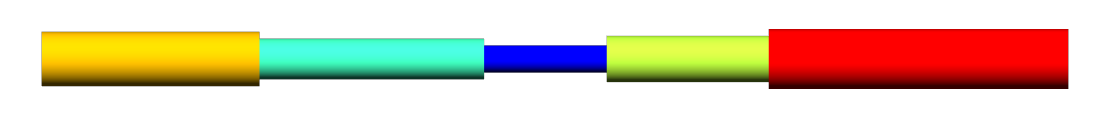} \caption{One-dimensional network of serially-connected cylindrical tubes.}
\label{Net1D1}
\end{figure}

Now we test the consequences of this modeling approach and compare the yield point obtained
dynamically from solving the pressure field to the one obtained statically from the threshold path
algorithms. In this regard, it is easy to verify that the dynamic \typ\ of the structure according
to this modeling strategy generally exceeds the sum of the \typ s of the individual tubes as given
by the threshold path algorithms. In Appendix \ref{AppB} we presented a mathematical proof for this
assertion for the case of a one-dimensional network consisting of an ensemble of serially-connected
tubes. We also demonstrate this by a simple example of such a network, shown in Figure \ref{Net1D1}
with data given in Table \ref{Net1DTable}, where we can simply verify that the sum of the \typ s is
450~Pa for a \ys\ value of 5~Pa, while the \typ\ for this value of \ys\ as obtained dynamically
from solving the pressure field is about 1664~Pa. Although a general proof for such a statement for
two-dimensional and three-dimensional networks are not available currently, we feel that the same
principles should apply. Anyway, the special case of one-dimensional networks is sufficient to
discredit the BRKMP claim that theses two thresholds are equal in general, as will be discussed
later in detail. Furthermore, all our simulations using the old and the new computational
frameworks with a diversity of two-dimensional and three dimensional networks produced dynamic
yield points that are generally larger than the static yield points.

\begin{table} [!h]
\caption{Data related to the network of serially-connected tubes shown in Figure \ref{Net1D1} where
$\Delta P_t$ stands for the \typ\ of the respective tube as given by Equation \ref{yieldCondition}
and the indices are related to the tubes in Figure \ref{Net1D1} from left to right. The \typ s
given in the fourth column are based on a \ys\ value of $\ysS=5$~Pa.} \label{Net1DTable}
\vspace{0.2cm} \centering
\begin{tabular}{c@{\hskip 3cm}c@{\hskip 3cm}c@{\hskip 3cm}c}
\hline
{\bf Tube Index} & {\bf $R$} & {\bf $L$} & {\bf $\Delta P_t$} \\
\hline
         1 &      0.020 &      0.160 &         80 \\

         2 &      0.015 &      0.165 &        110 \\

         3 &      0.010 &      0.090 &         90 \\

         4 &      0.017 &      0.119 &         70 \\

         5 &      0.022 &      0.220 &        100 \\
\hline
\end{tabular}
\end{table}

%XXXXXXXXXXXXXXXXXXXXXXX
\subsection{Solid-Like Approach}

According to this approach, the \ys\ materials prior to reaching their yield point are solid-like
substances. A reasonable modeling strategy for this approach is to determine the pressure field
from intuitive physical considerations as part of this approach. The most reasonable option for
modeling the pressure field in this case is to assume a linear pressure drop across the structure
and hence a constant pressure gradient. Any other model for the pressure field requires additional
justification. The propagation of pressure through solid materials may seem strange but it is
physically sound since pressure can propagate through solids as in the case of sound wave
transmission and reflection for instance. Although solids are normally assumed to be rigid, they
are not absolutely rigid since all solids, due to their atomic structure, have a certain degree of
elasticity; the \ys\ materials are not an exception as they obviously have such a property.

According to this modeling strategy and the associated assumption about the spatial definition of
the pressure field, the pressure field is determined as a function of the pressure drop across the
structure with no need for solving the balance equations as it can be obtained from pure geometric
considerations such as the proportionality of pressure to the distance from the inlet and outlet
boundaries. If a linear pressure drop is adopted to define the pressure field prior to yield then
no backtracking will occur. The \typ\ is found by increasing the pressure drop across the structure
gradually starting from a point known to be below the threshold yield point of the structure. A
test is then carried out at each pressure step to identify a possible connected route that spans
the structure from the inlet to the outlet with all its conduits being above their threshold yield
pressure. The minimum pressure that satisfies such a condition will be deemed as the \typ. At and
above this point, the flow in the mobilized part of the system should be determined by solving the
balance equations according to the presumed \ys\ theoretical fluid, such as Bingham, that is used
to model the flow following mobilization.

The state of the flow system, according to this solid-like scenario, is expected to change
radically on reaching the yield point and hence a very different pressure field may replace the
pre-mobilization pressure field. These two pressure fields could even be qualitatively different.
Furthermore, the system at the transition point may be unstable especially if complex non-Newtonian
rheological factors, such as history-dependent effects, are becoming involved in the post yield
processes. The occurrence of such instabilities is entirely realistic from the physical viewpoint
as such transitional instabilities are commonplace in physical systems, including fluid dynamics.
Mathematical models may also be characterized by such instabilities. Anyway, as indicated earlier
we are not concerned with these issues in the present paper which is limited in scope to the
identification of the \typ\ with a minor interest in other related issues. Any subsequent changes
in the dynamics of the system will not change the yield point which took place earlier as this is
part of the system history.

With regard to the consequences of this solid-like approach and the adopted modeling strategy, it
is easy to verify that the \typ\ of the flow system generally exceeds the sum of the \typ s of the
individual tubes on the threshold path as given by the threshold path algorithms. In Appendix
\ref{AppC} we presented a mathematical proof for this assertion for the case of a one-dimensional
network consisting of serially-connected tubes. We also demonstrate this by the simple example of
Figure \ref{Net1D1} and Table \ref{Net1DTable}, where we can easily verify that the sum of the \typ
s for a \ys\ value of 5~Pa is 450~Pa, while the \typ\ for this value of \ys\ as obtained
dynamically from inspecting the pressure field is about 754~Pa assuming the ensemble is straightly
aligned. The required threshold yield pressure gradient will be greater for a tortuous network
since the pressure gradient across the entire network will be multiplied by a sinusoidal factor to
obtain the component of the pressure gradient in the tube axial direction.

%XXXXXXXXXXXXXXXXXXXXXXXXXXXXXXXXXXXXXXXXXXXXXXXXXXXXXXXXXXXXXXXXX
\section{Pressure Regimes}\label{PressureRegimes}

\begin{figure}[!h]
\centering{}
\includegraphics
[scale=1] {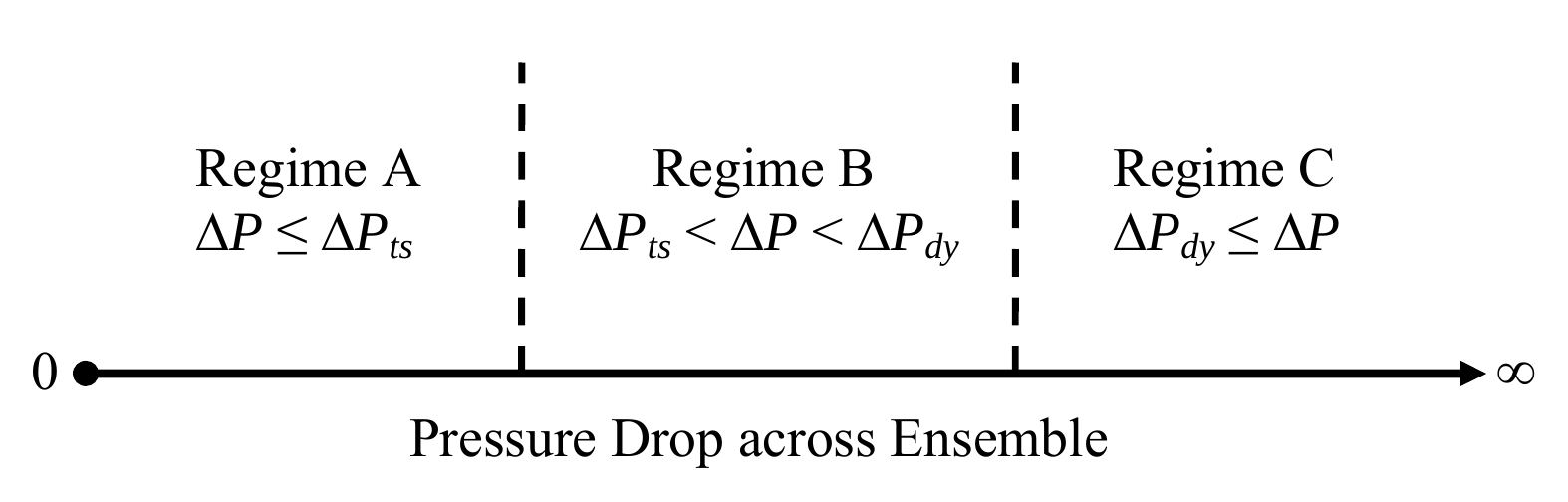}%
\caption{The three pressure regimes for an ensemble of interconnected conduits or a porous
structure, where $\Delta P_{ts}$ is the pressure drop of threshold sum and $\Delta P_{dy}$ is the
pressure drop of dynamic yield.} \label{RegimesFig}
\end{figure}

To clarify the situation for identifying the threshold yield and solidification points, we refer to
Figure \ref{RegimesFig} where we identified three mutually-exclusive pressure regimes related to
the magnitude of the applied pressure drop across an ensemble of interconnected conduits or a
porous structure. For simplicity, we assume the ensemble is a one-dimensional straightly-aligned
network like the one depicted in Figure \ref{Net1D1} although the classification and associated
arguments are valid in general for other types of networks and porous structures. These pressure
regimes are

\begin{itemize}

\item
Regime A where the pressure drop is less than or equal to the sum of the \typ s of the tubes in the
ensemble.

\item
Regime B where the pressure drop is larger than the threshold sum but less than the \typ\ as
identified by the dynamic argument based on solving or inspecting the pressure field.

\item
Regime C where the pressure drop is greater than or equal to the ensemble \typ\ according to the
dynamic argument.

\end{itemize}

We all agree that in regime A the ensemble is blocked because there is no way to split the pressure
drop to ensure simultaneous yield of all the tubes in the ensemble. This is correct whether we
applied the pressure drop gradually and continuously from above or from below or we applied the
pressure suddenly, as long as we start from a solid state point.

We also agree that in regime C the ensemble is open to the flow because whether we used the
rheological model of the \ys, like Bingham, or the pre-yield model, like \pois\ or solid state, the
pressure will split in both possibilities such that every tube in the ensemble will reach its yield
point. There is also no difference with regard to the sudden or gradual application of such a
pressure drop if we ignore, in the case of a sudden application, a possible brief transitional
stage during which the pressure adjusts itself to satisfy the requirement of one of the rheological
models and hence the system may still be blocked.

As for regime B, there are different scenarios that generally depend on the initial conditions and
the method of applying the pressure drop as outlined below

\begin{figure}[!h]
\centering{}
\includegraphics
[scale=0.75] {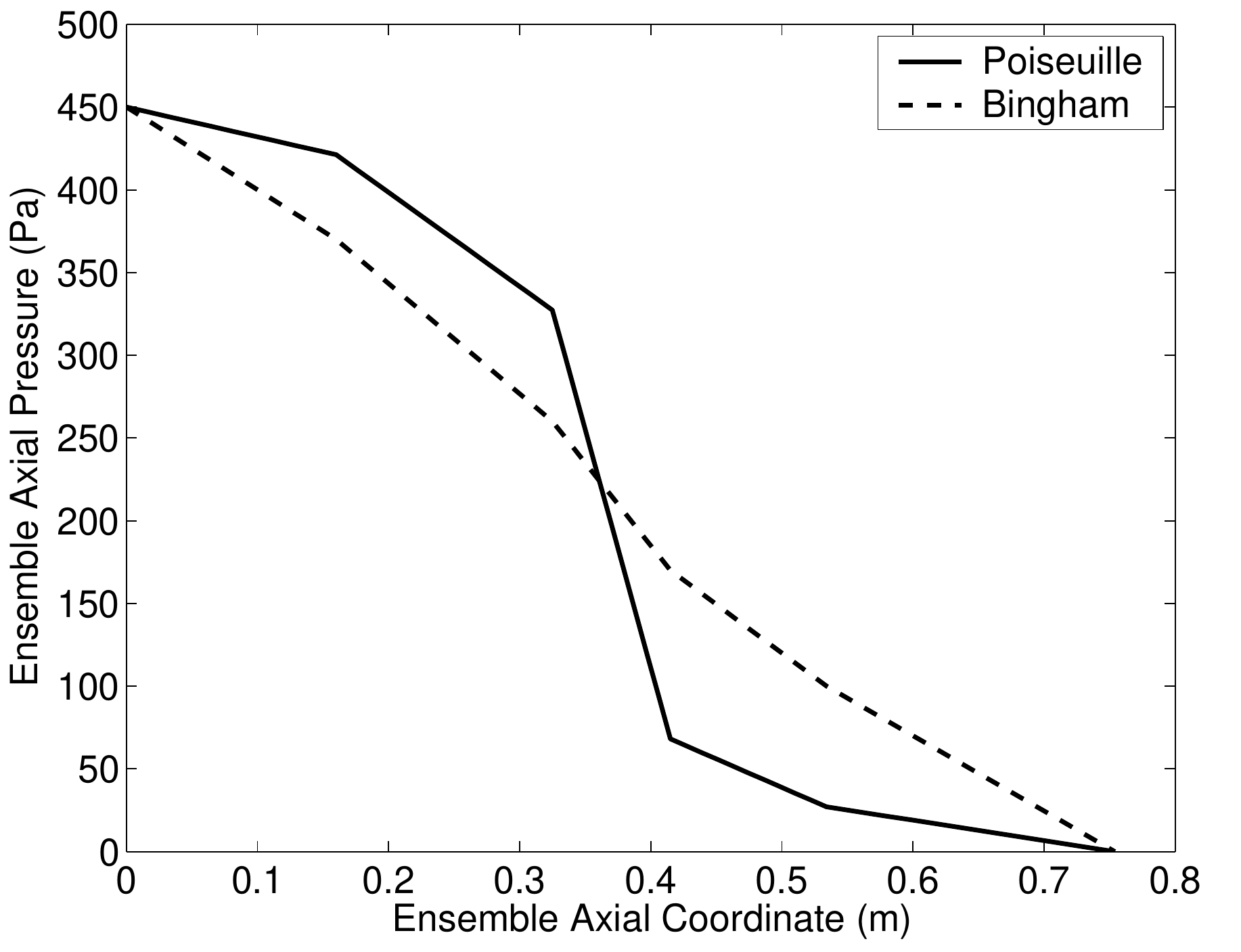}%
\caption{Axial pressure as a function of axial coordinate of the ensemble of Figure \ref{Net1D1}
for a \pois\ flow with a pressure drop of 450~Pa and a Bingham flow with a pressure drop of
450.1~Pa. The \ys\ of the Bingham fluid is assumed to be 5~Pa. The Bingham pressure field is
obtained by gradual decrease of the pressure drop starting from a high value belonging to regime C
where the system in known to have already reached its dynamic yield point. The flow rate for \pois\
is $Q_P\simeq2.26\times10^{-4}$~m$^3$.s$^{-1}$ assuming $\mu=0.05$~Pa.s, and for Bingham is
$Q_B\simeq2.72\times10^{-11}$~m$^3$.s$^{-1}$ assuming $C=0.05$~Pa.s.} \label{PoisBingFig}
\end{figure}

\begin{itemize}

\item
We should agree that if we start from a pressure drop in stage C where the system is flowing and
keep decreasing the pressure drop gradually and continuously then we should have a flow in stage B
as well, because the initial condition for the system requires the application of the \ys\
rheological flow model and hence all is needed is the satisfaction of the mass conservation
principle which is possible even in regime B according to the mathematical argument of Appendix
\ref{AppA}. This may be stated in a different way by saying that the smooth variation of the
pressure field inside the structure in response to a similar variation in the pressure drop across
it requires the continuity of the initial configuration of the pressure field which, qualitatively,
is that of a \ys\ rheology. The assumption of a sudden blockage on entering regime B implies a
sudden and non-continuous change in the pressure field configuration which is difficult to imagine
and justify physically. Now whether the system will be blocked or not on further decrease beyond
the lower limit of regime B is dependent on possible hysteresis, as discussed early.

\item
If we start from a pressure drop in regime A where the system is blocked and keep increasing the
pressure drop continuously then on exceeding the upper limit of this regime the system should be
still blocked because at the very edge of regime A we agree that the system is blocked since it is
subject to the pre-yield model and according to this model the pressure field is very different to
that required for a simultaneous yield of the tubes. It is difficult to imagine that an
infinitesimal increase in the pressure drop on passing the upper limit of regime A will change the
pressure field configuration suddenly and radically to the configuration required for a
simultaneous yield as a consequence of the supposed validity of the adopted \ys\ rheological model
on such a trivial transition. This is demonstrated in Figure \ref{PoisBingFig} where we compared
the pressure field of a \pois\ flow with a pressure drop of 450~Pa across the ensemble, which is
equal to the threshold sum, with the pressure field of a Bingham flow with a pressure drop of
450.1~Pa across the ensemble, which is just above the threshold sum. As can be seen, these two
pressure fields are very different. Such a strong dissimilarity will also be obtained for a
solid-like pre-yield approach. Therefore we think the most logical scenario is that the pressure
field will keep adjusting itself continuously and smoothly according to the rules of the pre-yield
model, whether fluid or solid, on a continuous increase of the pressure drop across the ensemble
all the way through regime B and hence the system will yield only when it enters regime C.

\item
Now if we apply a sudden pressure drop whose value belongs to regime B then the outcome in our view
is dependent on two factors: transitional instabilities and the previous pressure regime to which
the system was subject prior to the sudden change. Briefly, if the previous pressure is in regime A
then the most likely outcome is blockage, but transitional instabilities may lead to a pressure
distribution that opens all the tubes simultaneously and hence the system will continue flowing
because as soon as the system starts flowing, whether it is in regime B or C it should be subject
to the \ys\ rheological fluid model which can sustain a stable mass-conserving flow according to
the mathematical argument of Appendix \ref{AppA}. On the other hand if the previous pressure drop
belongs to regime C then the most likely scenario is mobilization although instabilities may lead
to blockage. Other static and dynamic factors, like hysteresis, should also play a role in these
scenarios. Other rheological aspects, especially history-dependent attributes, could also be
important in determining the transitional stage and the final outcome.

\end{itemize}

%XXXXXXXXXXXXXXXXXXXXXXXXXXXXXXXXXXXXXXXXXXXXXXXXXXXXXXXXXXXXXXXXX
\section{Discussion}\label{Discussion}

Now, the key question is why a \ys\ fluid model at a pressure just above the threshold sum produces
a mass-conserving consistent pressure field configuration with an open path while a solid-like or
Newtonian fluid models do not produce such an open path at such a pressure. In Appendix \ref{AppA}
we presented a mathematical argument to demonstrate why a mass-conserving consistent pressure field
can always be found for a pressure drop above the threshold sum of a \ys\ fluid-structure system.

In reality finding a mass-conserving consistent pressure field for a \ys\ fluid above the threshold
sum of an ensemble is not different to finding such a field for a \pois\ flow except that while for
the \pois\ flow the threshold pressure for mobilization is zero, for the \ys\ fluids the
mobilization threshold is the sum of the \typ s since this sum is the absolute minimum for any
possible mobilization assuming that it is split correctly to overcome the yield point for each tube
in the ensemble. As we always can find a mass-conserving consistent pressure field for \pois\ flow
above the zero pressure reference level, which seems self-evident although we believe it requires a
mathematical proof as outlined in Appendix \ref{AppA}, we can find such a pressure field for the
\ys\ fluids above the sum of thresholds for the same reason. The latter gives the illusion that
this is because the yield point is at the threshold sum whereas the reality is that justifying the
search for a consistent pressure field above the sum and below the dynamic yield point is only
justified if we assume that the system in regime B is in a yield state which can be justified in
the case of solidification process but not in the case of yield process.

The mathematical argument in Appendix \ref{AppA} reveals our main objection to the BRKMP reasoning
because what BRKMP do in their model is only to find a mass-conserving consistent pressure field
above the threshold sum point which can be trivially found. Finding such a pressure field gives the
impression that the actual yield point is at the threshold sum whereas in reality searching for
such a mass-conserving pressure field in the neighborhood of the threshold sum can only be
justified if we assume that the system is already in a yield state above the threshold sum point.

This reveals that all the derived results of BRKMP are in fact based on the very statement that
they are supposed to prove and hence they in fact use a circular argument. The matter of fact is
that they start from the assumption that the threshold yield pressure of the ensemble is detrmined
by the sum of the \typ s of the individual tubes; then all they need to take care of is mass
conservation above this limit.

Therefore, the BRKMP method, which is supposed to be for identifying the yield point, may be more
appropriate to use for identifying the blockage or solidification point because when they start
from a pressure point above the threshold sum point assuming the system is already in a fluid state
to which a rheological \ys\ fluid model, like Bingham, applies and keep lowering the pressure
gradually and continuously as can be concluded from their algorithm (refer e.g. to their equation
$\Delta P^{*}=\Delta P_m(1+\varepsilon)$), they will inevitably converge to the blockage point at
the threshold sum point.

However, in reality, due to hysteresis, the actual solidification point may be below the threshold
sum point as indicated previously. Such a hysteresis lagging is commonplace in polymeric and other
\ys\ systems, and hence it can delay the solidification to a pressure point below the value of
threshold sum when approaching the point from above. So, even if we assume that the yield point
from below is the one obtained by the threshold path algorithms it is not necessarily that the
solidification point from above is the same as the yield point. This of course implies that the
system during this pressure decreasing process will be subject to a different \ys\ rheological
model from the one that applies during the yield process or at least to the same rheological model
but with different parametric values. Briefly, the points of transition between the solid state and
fluid phase for \ys\ materials do not necessarily agree even in the bulk rheology regardless of the
extra reasons for this in the {\it in situ} rheology, which we are concerned with in this paper,
and whether the actual yield point for a network or a porous structure is at the threshold sum or
not. Experimental evidence has already shown that the two points usually do not agree. The reason
for hysteresis in general is the disturbance of the micro-molecular structure during the
deformation process in the fluid phase.

In fact even detecting the solidification point from above, ignoring the hysteresis issue, is only
legitimate if we start from a confirmed fluid state point as obtained from the dynamic yield
condition, i.e. from a point in regime C, and keep lowering the pressure all the way through regime
B until we reach the threshold sum point, because as explained early there is no ground in general
for assuming a yielded fluid state if we start in regime A or B. In our simulation experiments we
experienced exceptional convergence difficulties when we tried to start from pressure points in
regime B. We observed that the convergence was easier if we start from above the dynamic yield
point in regime C, which is unusual because convergence difficulties usually increase with higher
pressure boundary conditions. This is in complete agreement with the BRKMP observations about
convergence difficulties in these regimes (refer for example to their statement ``The traditional
Newton's method converges easily if the imposed pressure gradient is significantly higher than the
threshold pressure gradient...'') although BRKMP may offer a different explanation. The reason for
these difficulties in our view is the difficulty of finding a consistent pressure field of a
yielded system on starting from the given initial conditions based on the state in regime B.

In brief, we can challenge the underlying assumptions of BRKMP model that lead to such conclusions.
The key question that BRKMP should consider is why a pressure drop that is infinitesimally above
the sum of the threshold yield pressures applied across a serially-connected ensemble should
necessarily split, according to their yield scenario, such that the size of the pressure drop
across each tube is infinitesimally above its \typ, while at an infinitesimally lower pressure drop
across the ensemble (i.e. when the pressure drop was equal to the threshold sum) the pressure field
was very different as it was subject to a different rheological model. If we accept this
non-evident and controversial scenario, which BRKMP explicitly or implicitly present as a fact and
not just as an assumption or a possibility, then all is needed is to satisfy the mass conservation
principle which is a trivial thing to do as we demonstrated in Appendix \ref{AppA}. Therefore, it
is not surprising when BRKMP find that the threshold yield pressure as found from solving the
pressure field is identical to the value obtained from the threshold path algorithms because during
all the stages of stepping down on the pressure ladder they are using a \ys\ fluid model since they
assume, at least implicitly, that the yield-stress fluid has already yielded in regime B and hence
a physical flow that satisfies the mass-balance equation will be found inevitably. As soon as they
approach their `yield' point from above based on a fluid state assumption, they should converge to
a zero flow at the threshold sum and hence the two pressure values will necessarily agree.

%XXXXXXXXXXXXXXXXXXXXXXXXXXXXXXXXXXXXXXXXXXXXXXXXXXXXXXXXXXXXXXXXX
\section{Criticism}

We now address the main criticisms raised by BRKMP to our \ys\ modeling approach and the
conclusions that have been reached in SB. We also present some of our criticisms to the BRKMP model
as described in their paper.

One of the major criticisms directed to our model is that delaying the yield point beyond reaching
the threshold sum of the threshold path algorithms necessitates that some conduits have already
reached their yield points (refer to the BRKMP statement ``This condition requires that......and
not yield flow.'') and hence cannot be blocked as implied by our model which requires the yield
point to occur at a higher pressure belonging to regime C in a blatant violation to the conduit
yield condition as given by Equation \ref{yieldCondition}.

First, according to our model there is no ground for the application of the yield condition of
Equation \ref{yieldCondition} prior to reaching the yield point of the ensemble, because we are not
looking for the yield condition of a single tube but we are searching for the yield point of an
ensemble or porous structure. Before reaching the yield condition of the ensemble the material is
not considered a \ys\ fluid that is subject to the condition of Equation \ref{yieldCondition}, but
it is either a highly-viscous Newtonian fluid or a solid state material. Therefore before reaching
the dynamic yield point in regime C the flow system is assumed to be Newtonian or solid state and
hence no \ys\ does exist. Yield-stress model will take effect only on opening a spanning path that
sustains a tangible quantity of flow by reaching the ensemble dynamic yield point.

Second, the possibility that some conduits reach their yield point before the system reaches its
yield point occurs not only in the controversial B regime but even in regime A where we all agree
that the system in this regime cannot yield; whether we adopted a highly-viscous fluid approach or
a solid-like approach prior to yield. Even BRKMP who use, according to our understanding of their
model as indicated for example by the second part of their equation (1), a \pois\ model prior to
reaching the threshold sum should accept that some conduits will reach their yield condition as
given by Equation \ref{yieldCondition} in regime A. For example, the sum of threshold yield
pressures for the ensemble of Figure \ref{Net1D1} and Table \ref{Net1DTable} is 450~Pa for
$\ysS=5$~Pa. However, if we apply a pressure drop across the ensemble well below this sum then we
will find that some of the tubes have already reached their yield point assuming a \pois\ flow of a
highly-viscous fluid. For instance if we apply a pressure drop of 200~Pa then the pressure drop
across the third tube with a consistent pressure field of a \pois\ flow will be about 115~Pa which
is well above its \typ\ of 100~Pa assuming a \ys\ of 5~Pa. The solid-like approach also implies the
occurrence of such situations. So even according to the BRKMP modeling strategy such a `violation'
to the yield condition in some conduits is inevitable. The setting of the flow in these conduits to
zero, as BRKMP seem to suggest, is arbitrary and hence requires justification; moreover this
setting is a clear violation of the adopted \pois\ flow in these conduits in this regime which has
obvious consequences on the mass conservation balance. More discussion about this issue will be
presented later.

Third, and possibly the most important factor, is that pressure drop is a necessary but not a
sufficient condition for fluid flow. Two obvious examples are yield-stress fluids where no flow
occurs even with the presence of a pressure drop, and the second is a tube immersed vertically in a
body of water. The presumed solidity or fluidity with high viscosity in the first example cannot
change the argument which is based on the expectation of a tangible flow of a fluid phase prior to
yield as if it was a normal \pois\ fluid. For the second example the flow upwards will not happen
even with the presence of a pressure drop in the upward direction because it is balanced by another
force which is the force of gravity in this case. A conduit confined within a non mobilized
structure will not flow even if it reached its threshold yield pressure due to a similar balancing
force, that is the yield stress force of the surrounding structure which is essentially the same
force that prevents flow in a stand-alone tube filled with a \ys\ fluid and subjected to a pressure
drop below its yield point.

Another criticism to our model is the violation of local mass balance (refer to ``It is unknown why
Sochi (2010) obtains...found from search algorithms.'' in BRKMP). According to our model, the
system before reaching the dynamic yield point as obtained by solving the pressure field, assuming
a highly-viscous fluid approach, is subject to the \pois\ model and hence the mass is conserved
locally and globally. The illusion of a violation to the local mass balance arises from imagining
that the isolated throats will be automatically subject to the rheological model of the yielded
\ys\ fluid as soon as they reach their threshold yield condition of Equation \ref{yieldCondition}.
Mass balance violation can only occur if it is not accommodated in the flow model correctly, and
hence if the model dictates that the mobilization in the individual throats does not take place
automatically as soon as they reach their threshold yield point, but should also associate the flow
conductivity condition by being part of a yielded inlet-to-outlet spanning path, then no local or
global mass balance violation will occur.

In fact the mere distinction between local and global mass balance, as if they are two separate
conditions, is incorrect, because these two conditions are the same in essence due to the fact that
the global mass balance is based on the local mass balance of the individual interior pores. This
can be proved simply by stepping through the network from the inlet boundary to the outlet boundary
to verify that the total outflow must be equal to the total inflow if mass balance is respected on
each interior pore \cite{SochiTechnical1D2013}. In brief, local and global mass balance should be
satisfied if the process is modeled correctly using a consistent \ys\ flow model as described
early.

With regard to our criticism to the BRKMP investigation, in addition to the points that we already
made, we should first express our reservation about the graph theory proof. We have a strong
suspicion about the capability of the graph theory in principle to determine the outcome of a
physical process in such dynamic systems. All the graph theory, and any similar mathematical
apparatus, can do is to reproduce the pre-stated assumptions in a technical form with drawing some
logical conclusions from the given conditions. In fact the content of the given proof of graph
theory may not even be controversial as long as it is related to finding the threshold path from
static considerations. The important thing that really needs a proof is the underlying assumptions
and conditions which lead to these logical conclusions. The expected outcome of such processes in
such dynamic systems is therefore more logical to obtain from dynamic considerations based on the
physics of fluid mechanics.

We also observe that, unlike us, BRKMP do not have a model for the solid-like approach. The
assumptions of solid-like and highly-viscous fluid approaches are not just mathematical ideals but
they correspond to a physical reality, that is the \ys\ fluids should behave in one of these ways
or the other and hence for a complete modeling approach both possibilities should be considered. In
fact it is physically viable that even some \ys\ materials prior to yield could behave as
solid-like while others behave as highly-viscous fluids or a single \ys\ material behaves
differently under different physical conditions.

We also notice that what BRKMP describe as ``Close inspection'' in their statement ``Upon
convergence, some throats may appear open...(total flow into the network model equals flow out of
the model) is found.'' may not be sufficient to make such generalizations and hence if this is a
possible defect in the model it should be approached in a more formal, systematic and rigorous way
than a close inspection. Another point is that ``some throats may appear open'' just confirms what
we stated already about the inevitability of this situation even according to the BRKMP modeling
approach; the use of ``may appear'' to reduce the impact as if we are witnessing a real physical
process and not just a model that we created by our own hands does not make any good. We also do
not understand the supposed problem in mass conservation as if it is a matter of choice that we
need to take care of personally: simply if we set our model correctly and ensured that our code
does not contain serious bugs then mass conservation will be taken care of automatically by the
model and the code without need to worry about it and try to fix it through close inspection or
arbitrary blocking of some throats or any other means. The Newton-Raphson method as described by
BRKMP is sufficiently robust to conserve mass. Yes what should be worrying is a possible
inconsistency in the model itself where it is theoretically assumed that no throat can reach its
yield condition unless it is part of a connected path whereas the physics of the model requires
such a situation to occur, as discussed early. We also do not understand the role or the value of
this arbitrary discarding of the isolated throats apart from the possibility of adjusting the model
to make it look more consistent.

There are also some other issues which are relatively minor in the BRKMP assessment to SB. For
example, there are some misinterpretations of SB, e.g. the meaning of the dynamic effects which are
wrongly interpreted as of a viscous nature, like the meaning of this term in Chen \etal\
\cite{ChenRY2005}, whereas we clearly stated that it is related to the pressure field, and hence
some of the BRKMP arguments may not stand as they are. Also, the path of minimum pressure algorithm
is not an approximate method but it is rigorous within its validity domain. The algorithm is mainly
based on a linear pressure drop assumption prior to mobilization which is mostly relevant to the
solid-like approach for un-yielded yield-stress materials. Backtracking in such situations will not
be allowed because it does not occur for obvious physical reasons, as indicated early in this
paper. We also notice that there is a mention and even discussion of convergence problems with some
suggestions about how to overcome these problems and improve the rate of convergence by Sochi in
his thesis \cite{SochiThesis2007}. Another minor remark is that although a single open path at the
threshold pressure gradient is the most common possibility, multiple open paths are also possible.

There are possibly other limitations in the BRKMP \ys\ model which we suspect from reading the
method description in their paper, like possible inconsistency in the use of \pois\ and \ys\ models
in the pre and post yield regimes. However reaching a definite conclusion about these issues
requires further technical and coding details and more clarifications from Balhoff and coworkers,
which are not available to us. There are other controversial issues in BRKMP that can be challenged
but they are not related to \ys\ and hence are entirely out of the scope of the present paper.

%XXXXXXXXXXXXXXXXXXXXXXXXXXXXXXXXXXXXXXXXXXXXXXXXXXXXXXXXXXXXXXXXX
\section{Final Thoughts}

Finally, by what means we can verify which model is the `correct' one? Experimental evidence should
have the final word about most, but not all, of the previous issues which determine the validity
and applicability of any model. There are many limitations in the experimental procedures, their
results, interpretations and conclusions. Although we think that experimental evidence can in many
circumstances rule in or rule out some of the above mentioned models and scenarios, such as the
yield point of an ensemble and if it is at the threshold sum or at the dynamic yield point or may
even be at a different point, many other possibilities related to other phenomena, which are more
involved and less obvious, may not be possible to assess and reach a conclusion about
unequivocally. Some flow systems, like a highly complex non-Newtonian \ys\ fluid in a topologically
and geometrically complex porous medium, may be too complex to reach a definite conclusion about
their rheological behavior including their yield point due to the involvement of many intricate
factors. The quantitative difference between the two methods for determining the yield point, for
instance, may be absorbed in the overall error margin of the yield process. The difference between
the static yield point as determined by the threshold path algorithms and the dynamic yield point
as determined from the pressure field is obviously system dependent and hence the difference
between the two methods may not be sufficiently big in some cases for an unambiguous conclusion.

Another limitation of the experimental evidence is that in some circumstances although it can
endorse certain possibilities it cannot entirely rule out other possibilities. For example, there
is a possibility that there are different types of \ys\ fluids where each type has a distinctive
and different yield and rheological behavior. `Yield-stress' is a generic label that can
encapsulate many other physical attributes that characterize different yield-stress materials and
hence affect the overall behavior of the flow system including its yield and solidification points.
Although this may be difficult to imagine with regard to the yield and solidification points, it
could have an impact on other rheological attributes that, directly or indirectly, affect these
points.

We also should not rule out the possibility of \ys\ models, other than the ones that have already
been proposed in the literature including our own model, that could lead to a different and
possibly better prediction of the yield and solidification points. Regardless of any model, there
is also the possibility of a yield point different to the static and dynamic ones, most likely to
be in between, due to the involvement of other rheological and dynamic factors. The proposed yield
and solidification scenarios in the literature including the present paper are mostly based on a
pure logical reasoning with an implicit assumption of an ideal \ys\ material, and hence many
real-world physical factors are not fully incorporated in these models.

Regardless of all these controversial and uncontroversial issues, even if our criticism to the
BRKMP model is rejected, our \ys\ model as proposed in SB and elaborated in the present paper is at
least as valid as the BRKMP model from a pure modeling viewpoint based on the sensibility and
consistency criteria, as long as there is no independent and conclusive evidence, experimental or
otherwise, with or against one of these models or the other. In this paper we provided sufficient
clarifications and justifications to endorse the \ys\ modeling approach of SB regardless of the
validity or invalidity of any other model. We therefore believe that the BRKMP attempt to
disqualify the modeling approach of SB is void.

%XXXXXXXXXXXXXXXXXXXXXXXXXXXXXXXXXXXXXXXXXXXXXXXXXXXXXXXXXXXXXXXXX
\section{Conclusions}

The main conclusions reached in this study is the confirmation of the previous findings by Sochi
and Blunt \cite{SochiB2008} and Sochi \cite{SochiYield2010} with regard to the threshold yield
pressure of yield-stress materials residing in rigid networks of interconnected conduits or rigid
porous structures subject to a pressure field defined by two pressure boundary conditions. The
essence of the previous findings is that the dynamic yield point as obtained from solving or
inspecting the pressure field is generally higher than the static yield point found by the
threshold path algorithms. This is in a complete disagreement with Balhoff \etal\
\cite{BalhoffRKMP2012} who claimed to have proved that the threshold yield pressure obtained
dynamically is identical to the one found by the threshold path algorithms. We demonstrated that
what Balhoff \etal\ identified is more appropriate to be the solidification point on a gradual and
continuous lowering of the pressure drop starting from an established fluid state rather than the
yield point of a solid state material. However, even this could be challenged on the basis of the
inertial nature of complex fluids that may shift the solidification point to a lower pressure point
than the threshold sum.

\clearpage
%XXXXXXXXXXXXXXXXXXXXXXXXXXXXXXXXXXXXXXXXXXXXXXXXXXXXXXXXXXXXXXXXXXX
\phantomsection \addcontentsline{toc}{section}{Nomenclature} %
{\noindent \LARGE \bf Nomenclature} \vspace{0.5cm}

\begin{supertabular}{ll}
%
%$\sR$                 & strain rate (s$^{-1}$) \\
%$\sS$                 & stress (Pa) \\
$\mu$                   &   fluid dynamic viscosity (Pa.s) \\
$\ysS$                & yield-stress (Pa) \\
$\wsS$                & stress at tube wall (Pa) \\
\\
$C$                   & consistency coefficient in Bingham model (Pa.s) \\
$L$                   & tube length (m) \\
%$n$                   & flow behavior index \\
$P$                   & pressure (Pa) \\
$\Delta P$            & pressure drop (Pa) \\
$\Delta P_{dy}$      & pressure drop of dynamic yield (Pa) \\
$\Delta P_{t}$       & threshold pressure drop (Pa) \\
$\Delta P_{ts}$      & pressure drop of threshold sum (Pa) \\
$Q$                   & volumetric flow rate (m$^{3}$.s$^{-1}$) \\
$Q_B$                  & flow rate of Bingham model (m$^{3}$.s$^{-1}$) \\
$Q_P$                  & flow rate of \pois\ model (m$^{3}$.s$^{-1}$) \\
$R$                   & tube radius (m) \\
\end{supertabular}

\clearpage
\phantomsection \addcontentsline{toc}{section}{References} %
\bibliographystyle{unsrt}
%\bibliography{Bibl}

\appendix

\clearpage
%XXXXXXXXXXXXXXXXXXXXXXXXXXXXXXXXXXXXXXXXXXXXXXXXXXXXXXXXXXXXXXXXX
\section{Mass-Conserving Pressure Field}\label{AppA}

Assume we have a network consisting of $n$ serially-connected cylindrical tubes which generally
have different radii and lengths containing a yield-stress material. The sum of threshold yield
pressures of the individual tubes in such an ensemble, $\Delta P_{st}$, is given by

\begin{equation}
\Delta P_{st}=\sum_{i=1}^{n}\Delta P_{it}
\end{equation}
where $\Delta P_{it}$ is the threshold yield pressure of tube $i$. Now if we apply a total pressure
drop of $\Delta P_{st}$ across the ensemble and assume that this total pressure drop is divided
such that for each tube a pressure drop equal to its threshold yield pressure $\Delta P_{it}$
occurs across its length, then at this total pressure drop $\Delta P_{st}$ the flow in the system
is zero because all tubes are at their threshold yield pressure. Now let us assume that we added an
infinitesimal increase in the pressure drop, $\epsilon>0$, across the ensemble such that

\begin{equation}
\Delta P_{st}+\epsilon=\sum_{i=1}^{n}\Delta P_{it}+\sum_{i=1}^{n}\epsilon_{i}
\end{equation}
then there should be in principle a finite minute flow in each tube in the ensemble. Since the flow
rate is a continuous function of the pressure drop for each tube, then it is possible to adjust the
arbitrary and infinitesimal $\epsilon_{i}$ such that the flow rate in all tubes is the same within
a given error tolerance. For this same reason (i.e. the flow rate is a continuous function of
pressure drop for each tube) if we now increase $\epsilon$ infinitesimally, it should be possible
to divide this increase on the pressure drops of the individual tubes such that the flow rate in
all tubes is still the same within the given error tolerance. By doing this process of adding an
infinitesimal increase to the threshold sum $\Delta P_{st}$ repeatedly and dividing the increase on
the individual pressure drops appropriately as before, we can reach any pressure drop above the
threshold sum $\Delta P_{st}$ such that the flow rate in all tubes of the ensemble is the same
within the given error tolerance. This in essence is the same as finding a consistent pressure
field sustaining a total flow rate in the ensemble that conserves mass.

This mathematical argument can be applied to \pois\ flow as well to prove that it is always
possible to find a consistent pressure field that sustains a mass-conserving flow for any pressure
drop greater than zero across such ensembles. The argument can also be generalized to any other
characteristic flow above the threshold mobilization pressure of the ensemble for that particular
fluid.

\clearpage
%XXXXXXXXXXXXXXXXXXXXXXXXXXXXXXXXXXXXXXXXXXXXXXXXXXXXXXXXXXXXXXXXX
\section{Yield Condition for Fluid Approach}\label{AppB}

For a network consisting of serially-connected cylindrical tubes containing a yield-stress material
assumed to be a highly-viscous fluid prior to yield, the flow of the material can occur \emph{iff}
two conditions are simultaneously satisfied: (a) the mass is conserved throughout the ensemble and
(b) all tubes pass their threshold yield point simultaneously. In the following we show that for
such a network these two conditions require a threshold yield pressure that in general is greater
than the sum of the threshold yield pressures of the individual tubes. The threshold yield pressure
for a cylindrical tube is given by

\begin{equation}
\Delta P_{t}=\frac{2\tau_{o}L}{R}
\end{equation}
while the \pois\ flow, which is assumed to model the flow prior to mobilization, in such a tube is
given by

\begin{equation}
Q=\frac{\pi R^{4}\Delta P}{8\mu L}
\end{equation}
where $\Delta P_{t}$ is the threshold yield pressure of the tube, $\tau_{o}$ is the yield stress,
$L$ and $R$ are respectively the tube length and radius, $Q$ is the volumetric flow rate, $\mu$ is
the fluid dynamic viscosity, and $\Delta P$ is the pressure drop across the tube.

Now let us take the tube with the largest radius in this serially-connected network. Since the flow
through the ensemble will not occur unless this tube reaches its threshold yield point, then a
necessary condition for the flow to occur is that this tube reaches its yield point. We will see
later that this is also a sufficient condition for the flow to occur in the network assuming mass
conservation is satisfied. For this tube the flow rate at its threshold pressure is

\begin{equation}
Q_{b}=\frac{\pi R_{b}^{4}\Delta P_{bt}}{8\mu L_{b}}=\frac{\pi R_{b}^{4}}{8\mu
L_{b}}\frac{2\tau_{o}L_{b}}{R_{b}}=\frac{2\tau_{o}\pi R_{b}^{3}}{8\mu}
\end{equation}
where $b$ is an index marking this tube, and $\Delta P_{bt}$ is the threshold yield pressure of
this tube. Due to the mass conservation, this flow rate is the same for all the tubes in the
network, that is for any tube other than the one with the largest radius we have

\begin{equation}
Q_{i}=\frac{\pi R_{i}^{4}\Delta P_{i}}{8\mu L_{i}}=\frac{2\tau_{o}\pi
R_{b}^{3}}{8\mu}
\end{equation}
where $i$ is an index marking the other tube. On rearranging and simplifying we obtain

\begin{equation}
\Delta
P_{i}=\frac{2\tau_{o}L_{i}R_{b}^{3}}{R_{i}^{4}}=\frac{2\tau_{o}L_{i}}{R_{i}}\frac{R_{b}^{3}}{R_{i}^{3}}=\Delta
P_{it}\frac{R_{b}^{3}}{R_{i}^{3}}
\end{equation}
where $\Delta P_{it}$ is the threshold yield pressure of tube $i$. Now

\begin{equation}\label{SufEq1}
R_{b}\ge R_{i}\,\,\,\,\,\Rightarrow\,\,\,\,\,\Delta P_{i}\ge\Delta P_{it}
\end{equation}
and hence the sum of the actual pressures across the individual tubes for such a flow assuming mass
conservation is greater than or equal to the sum of threshold yield pressures of the individual
tubes. The equality holds only when all the tubes in the network have the same radii. The condition
in Equation \ref{SufEq1} also explains why reaching the threshold yield pressure for the tube with
the maximum radius is not only a necessary condition but is also a sufficient condition for the
flow to occur assuming mass conservation, as indicated earlier.

\clearpage
%XXXXXXXXXXXXXXXXXXXXXXXXXXXXXXXXXXXXXXXXXXXXXXXXXXXXXXXXXXXXXXXXX
\section{Yield Condition for Solid Approach}\label{AppC}

For a network consisting of serially-connected and straightly-aligned cylindrical tubes containing
a yield-stress material which is assumed to be solid-like prior to its mobilization, the flow of
the material can occur \emph{iff} all tubes pass their threshold yield point simultaneously. In the
following we show that for such a network this condition requires a threshold yield pressure that
in general is greater than the sum of the threshold yield pressures of the individual tubes
assuming a linear pressure drop which is equivalent to a constant pressure gradient.

We take the tube with the smallest radius in the network. Since the flow through the ensemble will
not occur unless this tube reaches its threshold yield point, then a necessary condition for the
flow to occur is that this tube reaches its yield point. We will see later that this is also a
sufficient condition for the flow to occur in the network. For this tube, indexed by $b$, the
threshold yield pressure is given by

\begin{equation}
\Delta P_{bt}=\frac{2\tau_{o}L_{b}}{R_{b}}
\end{equation}
and hence the constant pressure gradient across the entire network when tube $b$ is at its
threshold yield pressure will be

\begin{equation}
\nabla P=\frac{\Delta P_{bt}}{L_{b}}=\frac{2\tau_{o}}{R_{b}}
\end{equation}
Now since the pressure drop is assumed linear with respect to the network total length, the
pressure drop across any other tube in the network, indexed by $i$, will be

\begin{equation}
\Delta P_{i}=\nabla PL_{i}=\frac{2\tau_{o}L_{i}}{R_{b}}
\end{equation}

Now since $R_{i}\ge R_{b}$ we have

\begin{equation}\label{SufEq2}
\Delta P_{it}=\frac{2\tau_{o}L_{i}}{R_{i}}\,\,\,\,\,\Rightarrow\,\,\,\,\,\Delta P_{i}\ge\Delta
P_{it}
\end{equation}

Hence, at the yield point of the ensemble the sum of the actual pressures across the individual
tubes in such a network based on the solid-like assumption with a constant pressure gradient across
the network is greater than or equal to the sum of threshold yield pressures of the individual
tubes. The equality holds only when all the tubes in the network have the same radii. The condition
in Equation \ref{SufEq2} also explains why passing the threshold yield pressure for the tube with
the minimum radius is not only a necessary condition but is also a sufficient condition for the
flow in the network to occur.

\end{document}